\documentclass[twocolumn,prb,superscriptaddress,showpacs]{revtex4-1}
\usepackage{amsmath}    
\usepackage{graphicx}   
\usepackage{verbatim}   
\usepackage{color}      
\usepackage{subfigure}  
\usepackage[pdffitwindow,colorlinks,citecolor={red},linkcolor={blue}]{hyperref}
\usepackage[dvipsnames]{xcolor}
\usepackage{multirow}
\usepackage{soul}
\usepackage{mathbbol}
\usepackage{bm}

\usepackage{fancyvrb}

\begin{document}

\title{Pure spin current injection of single-layer monochalcogenides}

\author{Bernardo S. Mendoza}
\email[E-mail: ]{bms@cio.mx}
\affiliation{Centro de Investigaciones en \'Optica, Le\'on,
Guanajuato 37150, M\'exico}

\author{Lucila Ju\'arez-Reyes}
\affiliation{Centro de Investigaciones en \'Optica, Le\'on,
Guanajuato 37150, M\'exico}

\author{Benjamin M. Fregoso}
\affiliation{Department of Physics, Kent State University, Kent, Ohio 44242,
USA}

\date{\today}

\begin{abstract}
We compute the spectrum of pure spin current injection in ferroelectric
single-layer SnS, SnSe, GeS, and GeSe. The formalism takes into account the
coherent spin dynamics of optically excited conduction states split in energy
by spin orbit coupling. The velocity of spins is calculated as a function of
incoming photon energy and angle of linearly polarized light within a full
electronic band structure scheme using density functional theory. We find peak
speeds of 250, 210, 180 and 154 Km/s for SnS, SnSe, GeS and GeSe, respectively
which are an order of magnitude larger than those found in bulk semiconductors,
e.g., CdSe and GaAs. Interestingly, the spin velocity is independent of the
direction of polarization of light in a range of photon energies. Our results
demonstrate that single-layer SnS, SnSe, GeS and GeSe are candidates to
produce on demand spin-velocity injection for spintronics applications.
\end{abstract}


\maketitle

\section{Introduction}
\label{sec:introduction}
There is increasing interest in attaining precise control of the spin of
electrons at mesoscopic scales because it could lead to novel quantum
computation
platforms.\cite{wolfSC04,awschalomNP2007,fabianAPS07,awschalomSSBM13}
 Pure spin
current (PSC), i.e., spin current with no associated charge current, could lead
to more efficient quantum devices because a PSC does not produce joule heating.
A PSC can be realized in the spin Hall effect,\cite{sinovaPRB04} one photon
absorption of light,\cite{bhatPRL05, nastosPRB07,zapataPRB17} or interference of two optical
beams.\cite{bhatPRL00, najmaiePRB03}

In contrast to the well-known spin Hall effect which occurs in the ground state
of the material, optically induced PSC occurs in excited states of the
material. Linearly polarized light injects carriers symmetrically into $\pm \bm{k}$
conduction states in the Brillouin zone (BZ). Yet, the velocity
and spin operators are odd under time reversal symmetry and hence there is no net charge current
\cite{alvaradoPRL85,schmiedeskampPRL88,bhatPRL05} or spin density after
summation over all the BZ.~\cite{Fregoso2019} However, the spin current is even
under time reversal and hence it does not vanish after summation over
all the BZ. The optical PSC will manifests as a spatial separation of spin-up spin-down
components or electrons. Optical PSC has been measured by pump-prove techniques
in gallium arsenide (GaAs),\cite{zhaoPRL2006, stevensPRL03} aluminum-gallium
arsenide (AlGaAs)\cite{stevensPRL03}, Co$_2$FeSi.\cite{kimuraNGPAM12}
and ZnSe.\cite{hubnerPRL03}

A major problem in the field of spintronics is the very short spin relaxation time
scales\cite{murakamiSc03} which makes it difficult to maintain coherence required for a spin
current.~\cite{majumdarAPL06,dattaAPL90,gotteNat16,pershinPRB08}  For this reason we turn our attention to
novel two-dimensional (2D) materials. 2D materials represent the ultimate
scaling in thickness with mechanical, optical, and electronic properties that
are unique relative to their bulk counterparts. For example, single-layer or 2D
group-IV mono-chalcogenides GeS, GeSe, SnS, and SnSe are actively being
investigated,\cite{liJACS13,singhAPL14,wangNM16,ramasamyJMC16,wuNL16,kamalPRB16,
guoJAP17,xinJPC16,hanakataPRB16,rangelPRL17,mendozaPRB20} due to their band
gaps and large carrier mobilities suitable for optoelectronics. Monochalcogenides are
centrosymmetric in the bulk but lack inversion symmetry in single-layer
form, thus allowing the generation of nonlinear effects such as optical PSC.

In this paper we show that 2D ferroelectric GeS, GeSe, SnS, and SnSe exhibit
large optical PSC as measured by the spin velocity injection (SVI) to be
defined below. We compute the spectrum of SVI as a function of photon energy
and angle of linearly polarized light. We consider the case of the particles moving
on the plane of the slab and spin pointing out of the slab. Monolayer SnS was
recently experimentally realized,\cite{Higashitarumizu2020} and hence our theoretical results have
direct experimental relevance for this material.

This paper is organized as follows. In Section \ref{sec:theory},  we present
the  PSC theoretical formalism, showing the main expressions used  to
numerically implement  PSC and SVI calculations. In Section \ref{sec:struc} we
describe the 2D monochalcogenide SnS, SnSe, GeS and GeSe structures and  in
Section \ref{sec:results}, we describe the results corresponding to the SVI
spectra for the chosen structures. Finally, we summarize our findings in
Section \ref{sec:conclusions}.

\section{Theory}
\label{sec:theory}
In this section we follow the formalism of Ref.~\onlinecite{zapataPRB17} and
present only the main theoretical results. We consider free Bloch electrons
subject to an external homogeneous electric field

\begin{align}
\bm{E}(t) = \bm{E}(\omega) e^{-i\omega t} + c.c..
\label{z.3}
\end{align}
The main idea is to extract an effective dynamics of spinors states
$c'c$ in the conduction bands. The key approximations are (i) hole spins do not
contribute to the current, (ii) the energy split between spinors $c,c'$ is
small compared with the energy difference between either of the spinors energy and any
valence band, and (iii) the frequency of the optical field is much larger than the
energy split between the spinors. The equation of motion of the effective density matrix
$\rho(\bm{k};t)$ is,\cite{zapataPRB17,nastosPRB07}
\begin{align}
\frac{\partial \rho_{cc'}(\bm{k};t)}{\partial t} &=
\frac{e^2}{i\hbar^2}E^{a}(\omega)E^{b}(-\omega)\sum_{v} r_{cv}^{a}(\bm{k}) r_{vc'}^{b}(\bm{k})
\\\nonumber
&\times\bigg(\frac{1}{\omega-\omega_{c'v}(\bm{k})-i 0^{+}}
 - \frac{1}{\omega-\omega_{cv}(\bm{k})+i 0^{+}} \bigg)
,
\end{align}
with $\hbar\omega_{n}(\bm{k})$
the energy of the electronic band $n$ at  point $\bm{k}$ in the
irreducible Brillouin zone (IBZ),  $|n\bm{k}\rangle$ is the Bloch state,
$\omega_{nm}(\bm{k})\equiv\omega_{n}(\bm{k})-\omega_{m}(\bm{k})$,
$H_0$ is the free Bloch Hamiltonian which
includes the spin-orbit coupling (SOC) and
$H_{0}|n\bm{k}\rangle = \hbar\omega_{n}(\bm{k})|n\bm{k}\rangle$.
Here $c,c'$ are time-reversed spinor states which are split in energy by
SOC.
The denominators clearly indicate the resonance coming from
the absorption of a photon with energy $\hbar\omega$, as the electron goes from
the valence state $v$ to either of the quasidegenerate states $c$ or $c'$.
Latin superscripts indicate Cartesian coordinates
and are summed over if repeated.

In this article, we focus on typical ultrafast experiments,\cite{stevensPRL03,hubnerPRL03}
for which one can ignore spin relaxation. \cite{bhatPRL05}
Thus, we neglect the
precession of the spins due to the spin splitting of the
bands as the precession period is long
compared to the momentum scattering time.\cite{dyakonov84}
Therefore, the spin current operator is  given by
\begin{align}
\hat K^{ab}(\bm{k})
 = \frac{1}{2}( \hat v^{a}(\bm{k}) \hat S^{b}(\bm{k}) +\hat  S^{b}(\bm{k}) \hat v^{a}(\bm{k})),
\label{z.1}
\end{align}
where $\hat{\bm{v}}=[\hat{\bm{r}},H_0]/i\hbar$  is the velocity
operator,  $\hat{\bm{r}}$ is the position operator,
$\hat{\bm{S}}=\hbar \hat{\bm{\sigma}}/2$ is the spin operator with $\hat{\bm{\sigma}}$ the Pauli spin
matrices, and
we allowed for the fact that in general
$\hat{\bm{v}}$ and $\hat{\bm{S}}$ do not commute.

We compute the average of
 the spin current injection tensor $\dot{\bm{K}}(\bm{k})$  as
\begin{align}
\dot{K}^{ab}(\omega) = \int \frac{d^{3}k}{8 \pi^{3}}
\sum_{cc'}
\frac{\partial \rho_{cc'}(\bm{k};t)}{\partial t} K^{ab}_{c'c}(\bm{k})
,
\label{eq:dodt}
\end{align}
where
we used the closure relationship
$|n\bm{k}\rangle\langle n\bm{k}| =1$,
and the integral is over the IBZ.
Then, we write
\begin{align}
\dot{K}^{ab}(\omega) = \mu^{abcd}(\omega) E^{c}(\omega) E^{d}(-\omega),
\label{eq:dotk}
\end{align}
where
\begin{align}
\mu^{abcd}(\omega) &= \frac{\pi e^{2}}{\hbar^{2}}
\int \frac{d^{3}k}{8 \pi^{3}}
\sum'_{vc c'}
K^{ab}_{cc'}(\bm{k}) r^{c}_{cv}(\bm{k}) r^{d}_{vc}(\bm{k})
\nonumber\\
&\times[\delta(\omega-\omega_{c'v}(\bm{k})) +  \delta(\omega-\omega_{cv}(\bm{k})) ]
,
\label{eq:mu}
\end{align}
is the tensorial response function that characterises the PSC.
The prime symbol  the sum means that $c$ and $c'$ are quasi-degenerate conduction
states, and the sum only covers these states.
Finally,
\begin{align}
K^{ab}_{cc'}(\bm{k}) = \frac{1}{2} \sum_{l=v,c} (v^{a}_{cl}(\bm{k}) S^{b}_{lc'}(\bm{k}) +S^{b}_{cl}(\bm{k})
v^{a}_{lc'}(\bm{k})),
\label{eq:velspimatelem}
\end{align}
are the matrix elements of Eq.~\eqref{z.1}.

An important point is that both charge current and net spin density
vanish and hence the current is a PSC. For linear polarization,
the momentum distribution of carriers is even in $\pm \bm{k}$. Assuming time
reversal symmetry, the velocity is odd in $\bm{k}$ and hence charge (injection)
current vanishes after summation over the whole BZ.~\cite{Fregoso2019}
Similarly, the spin matrix elements are odd in $\bm{k}$ and so the net spin
density vanishes. The spin current \ref{z.1} however, is even in $\bm{k}$ and
 does not vanish under linearly polarized light. Note that PCS with linearly
polarized light vanishes in the absence of SOC and that other definitions of the spin
operator,\cite{shiPRL06} are also even under time reversal symmetry.

\subsection{Spin velocity}
\label{sec:spin_vel}
To quantify the speed of the particles we define an effective charge velocity as\cite{bhatPRL05,zapataPRB17}
\begin{align}
\frac{\hbar}{2} \dot{n}(\omega) \mathbb{v}^{ab}(\omega) \equiv
  \dot{K}^{ab}(\omega).
\label{eq:vab-w}
\end{align}
$\mathbb{v}^{ab}$ gives the velocity of electrons along Cartesian direction $a$ with spin
polarized along Cartesian direction $b$.  The carrier injection rate $\dot n(\omega)$ is\cite{nastosPRB07}
\begin{align}
\dot{n}(\omega) = \xi^{ab}(\omega) E^{c}(\omega) E^{d}(-\omega),
\label{eq:dotn}
\end{align}
where
\begin{align}
\xi^{ab}(\omega)  = \frac{2\pi e^{2}}{\hbar^{2} V}
\int \frac{d^{3}k}{8 \pi^{3}} \sum_{vc}
r^{\mathrm{a}}_{vc}(\bm{k}) r^{\mathrm{b}}_{cv}(\bm{k})\delta(\omega-\omega_{cv}(\bm{k})),
\label{eq:xi}
\end{align}
is related to the imaginary part of the linear optical response tensor by
$\mathrm{Im} [\epsilon^{ab}(\omega)] = 2\pi\epsilon_0\hbar\xi^{ab}(\omega)$.

The function $\mathbb{v}^{ab}(\omega)$ allows us to quantify two very important aspects
of PSC. First, we can fix the spin along direction $b$ and calculate the
resulting electron velocity. Second, we can fix the velocity of the electron
along $a$ and study the resulting direction along which the spin is
polarized. In this article we restrict to the first case of the spin polarized
along $z$ to take the advantage of the 2D nature of the chosen
monochalcogenides, where the spin would then be polarized perpendicular to the
plane of the structures, as seen in Fig.~\ref{fig1}.  To this end, we  use an
incoming linearly polarized  light at normal incidence, and use the  direction
of the polarized  electric field to control $\mathbb{v}^{ab}(\omega)$. Indeed, writing
$\bm{E}(\omega) = E_0(\omega)(\cos\alpha\,\hat{\mathbf
x}+\sin\alpha\,\hat{\bm{y}})$, where $\alpha$ is the polarization angle with
respect to $x$, we obtain from Eq. \ref{eq:vab-w} that
\begin{align}
\mathbb{v}^{xz}(\omega,\alpha) =\frac{2}{\hbar} \frac{ \mu^{xzxy}(\omega)\sin 2\alpha }
{\xi^{xx}(\omega)\cos^2\alpha +\xi^{yy}(\omega)\sin^2\alpha},
\label{e.z21}
\end{align}
and
\begin{align}
\mathbb{v}^{yz}(\omega,\alpha)=\frac{2}{\hbar} \frac{\mu^{yzxx}(\omega)\cos^2\alpha +
\mu^{yzyy}(\omega)\sin^2\alpha } {\xi^{xx}(\omega)\cos^2\alpha +\xi^{yy}(\omega)\sin^2\alpha},
\label{e.z22}
\end{align}
where $\mu^{xzxx}(\omega)=\mu^{xzyy}(\omega)=\mu^{yzxy}(\omega)=0$ due to the  $mm2$ point group
symmetry of the chosen 2D monochalcogenides; we remark that only the
$\mu^{azbc}(\omega)$ are involved for the spin polarized along $z$. The speed of the
injected spin along $z$ is given by
\begin{align}
\mathbb{v}^{z}(\omega,\alpha) =\sqrt{(\mathbb{v}^{xz}(\omega,\alpha))^2 +(\mathbb{v}^{yz}(\omega,\alpha))^2},
\label{e.z23}
\end{align}
that makes an angle $\theta^{z}(\omega,\alpha)$ with respect to $x$ given by
\begin{align}
\theta^{z}(\omega,\alpha) =\tan^{-1}\left(\frac{\mathbb{v}^{yz}(\omega,\alpha)}{\mathbb{v}^{xz}(\omega,\alpha)}\right).
\label{e.z24}
\end{align}
We see that $\mathbb{v}^{xz}(\omega,\alpha)$ and $\mathbb{v}^{yz}(\omega,\alpha)$  have a
mirror plane at $\alpha=90^\circ$ and a period of $180^\circ$.  We define the
figure of merit for the speed of the injected spin along $z$ by

\begin{align}
v^{z}_a(\omega)=\mathbb{v}^{az}(\omega,\alpha=\pi/4),\quad a=x,y .
\label{e.z25}
\end{align}

\section{Structures}
\label{sec:struc}
\begin{figure}[]
\centering
\includegraphics[width=\columnwidth]{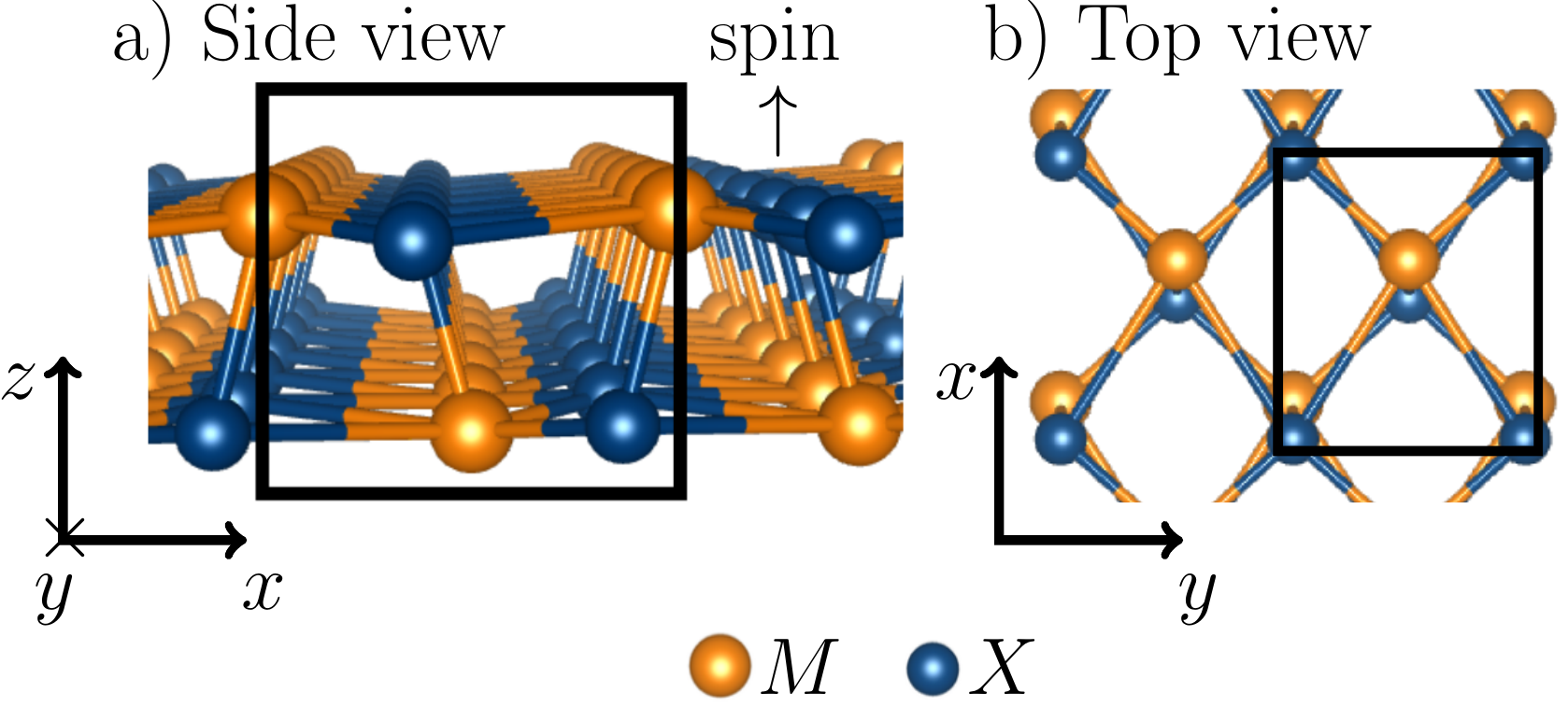}
\caption{(color online) The crystal structure of single-layer group-IV
chalcogenides MX, where M=Ge, Sn, and X=S, Se. In a) we show the 3D view of the
single-layer and in b) the projections of the single-layer crystal on the
Cartesian axes. The black rectangles denotes the unit cell. The upward arrow
denotes the $z$ direction of the spin.}
\label{fig1}
\end{figure}
Bulk monochalcogenide crystals MX (M = Ge, Sn and X= S, Se) are orthorhombic
with point group $mmm$ and space group $Pnma$ (No. 62).
They  consist of van der Waals-bonded double layers of metal  monochalchogenide
atoms in an armchair arrangement. The space group of  the bulk crystal contains
eight symmetries including a center of inversion which prevents pure spin
current (PSC). Upon exfoliation, the resulting single ``double layer''
primitive cell has four atoms as  seen in Fig.~\ref{fig1}, with the layers
chosen perpendicular to the $z$ axis. The single-layer structure has four
symmetries, including a two fold rotation with respect to $x$ (plus
translation), 2[001] + (1/2,0,1/2). In addition, the 2D system has two mirror
symmetries with respect to $z$ and $y$, (1/2,1/2,1/2) and $m$[010] + (0,1/2,0),
that leads to an $mm2$ point group, which determines the nonzero components of
the optical response tensors, like $\mu^{abcd}(\omega)$ and
$\xi^{ab}(\omega)$. The atomic slab widths are 2.84, 2.73, 2.56 and 2.61 {\AA}
for SnS, SnSe, Ges, GeSe respectively.

\section{Numerical Method}
\label{sec:numerical}
We calculated the self-consistent ground state and the Kohn-Sham states within
density functional theory in the independent-particle scheme within the local
density approximation (DFT-LDA), with a plane wave basis using the
ABINIT.\cite{gonzeCPC09}  The
Hartwigsen-Goedecker-Hutter (HGH) relativistic separable dual-space Gaussian
pseudopotentials,\cite{hartwigsenPRB98} are used to include the spin-orbit
interaction needed to calculate $K^{\mathrm{ab}}_{cc'}$ from Eq.
\ref{eq:velspimatelem}.
The convergence parameters for the calculations,  for all the structures  are a
cutoff energy of 30\,Ha, resulting in LDA band gaps of 1.35, 0.80, 1.82 and
1.05 eV for SnS, SnSe, GeS, GeSn, respectively. The TINIBA code,\cite{tiniba}
was used to calculate the response functions for which 4356 $\bm{k}$ points in
the IBZ were used to integrate $\mu^{abcd}(\omega)$ and $\xi^{ab}(\omega)$
using the linearized analytic tetrahedron method.\cite{nastosPRB07} We
neglect the anomalous velocity term
$\hbar(\boldsymbol{\sigma}\times\nabla V)/4m^2c^2$, where $V$ is the crystal
potential, in $\hat{\mathbf v}$ of Eq.~\ref{z.1}, as this term is known to give
a small contribution to PSC.\cite{bhatPRL05} Therefore,
$[\hat{\bm{v}},\hat{\bm{S}}]=0$, and Eq.~\ref{z.1} reduces to $\hat K^{ab}=\hat
v^{a} \hat S^{b}=\hat S^{b} \hat v^{a}$. Finally, the prime in the sum of
Eq.~\ref{eq:mu} is restricted to quasi-degenerated conduction bands $c$ and
$c'$ that are closer than 30 meV to each other, which is both the typical laser
pulse energy width and the thermal  room-temperature energy level
broadening.\cite{nastosPRB07} We include 20 valence and 40 conduction bands,
which accounts for all allowed transitions up to 6 eV. To model the slabs we
use supercells of 20 {\AA} along $z$, which corresponds to vacuum larger than
$17$ {\AA}, and renormalize our results to the atomic slab widths mentioned in
the previous section, thus removing the vacuum as it must.

\section{Results}
\label{sec:results}
We present results for monolayer SnS, which was  experimentally
realized recently,\cite{Higashitarumizu2020} as a representative example of the four
monochalcogenides studied. In the appendix we shows the results for SnSe, GeS
and GeSe.

\begin{figure}[]
\centering
\includegraphics[scale=.7]{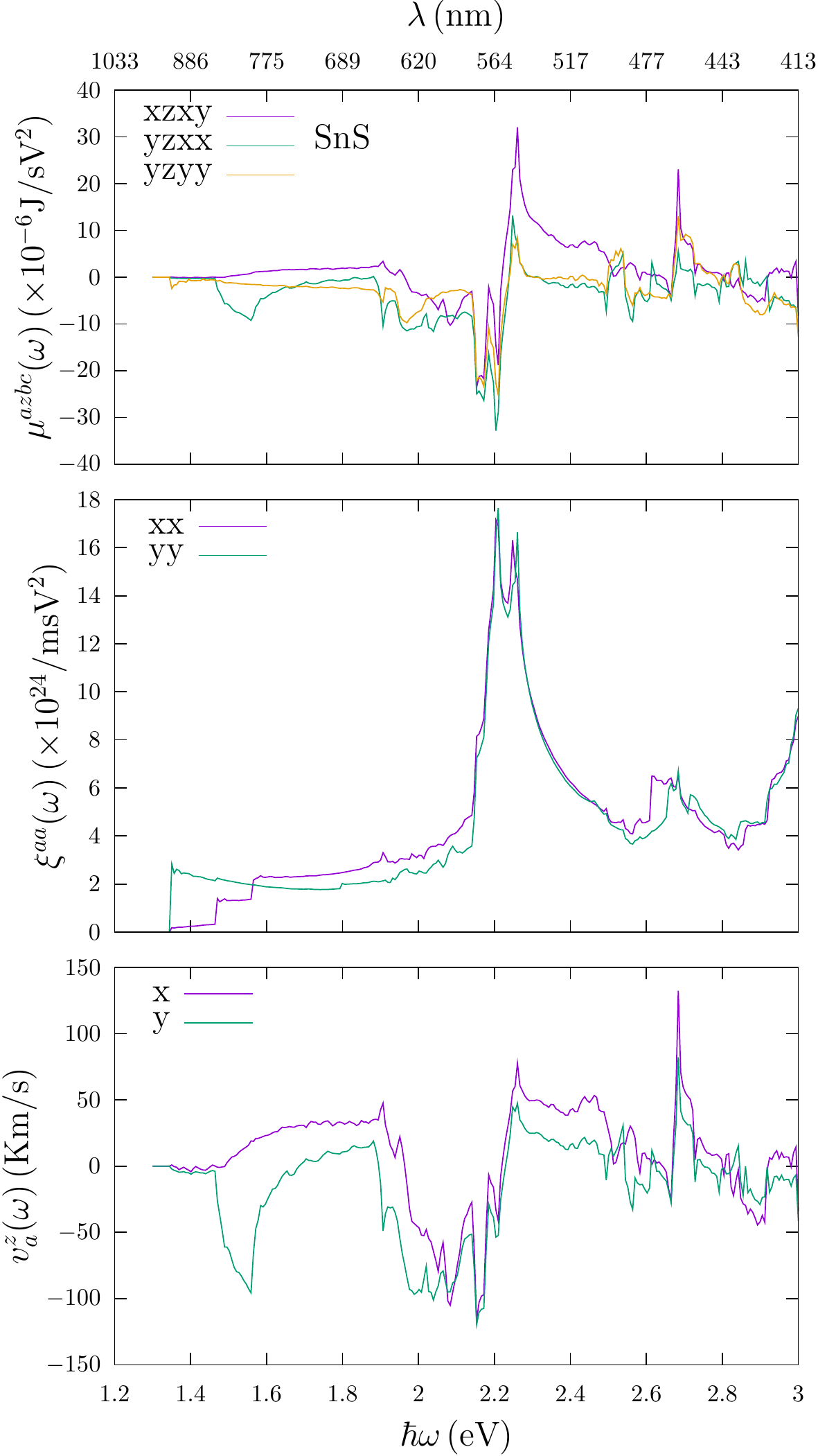}
\caption{(Color Online) $\mu^{azbc}(\omega)$, $\xi^{aa}(\omega)$, and $v^z_{(x,y)}(\omega)$ of
Eqs.~\eqref{eq:mu}, \eqref{eq:xi}, and \eqref{e.z25} respectively, vs
$\hbar\omega$ for SnS. }
\label{fig:mu-chi-vaz-sns}
\end{figure}

In Fig.~\ref{fig:mu-chi-vaz-sns}, we show  as a function of $\hbar\omega$,
$\mu^{azbc}(\omega)$, $\xi^{aa}(\omega)$, and $v^z_{x,y}(\omega)$  of Eqs.~\eqref{eq:mu},
\eqref{eq:xi}, and Eq. \eqref{e.z25}, respectively. The latter  gives the
figure of merit, for SnS for the SVI (spin-velocity injection), and remark that
$\xi^{\mathrm{aa}}(\omega)$ is a positive definite function. We only show the results
in the visible  range, where
there are ample sources of devices to produce light of the required energy or
its corresponding wavelength $\lambda
(\mathrm{nm})=1240/\hbar\omega(\mathrm{eV)}$. We see  that $v^z_{x,y}(\omega)$, which
is the central result of this article, has a rich structure as a function of
$\hbar\omega$, and more importantly, reaches values around $\pm 100$ Km/s,
albeit not necessarily  at the same energies; similar values of the SVI were
predicted for hydrogenated graphene.\cite{zapataPRB17} Right at the energy gap,
we see the onset of $\mu^{azbc}(\omega)$ and
$\xi^{aa}(\omega)$, and correspondingly of $v^z_{x,y}(\omega)$. There are three energy regions
where $v^z_{x,y}(\omega)$ are large, to wit, just below 1.6\, eV in the Red subrange of
the visible, between 2.0 and 2.2\,eV covering Orange, Yellow and Green colors,
and around 2.7\,eV in the Blue.
The behaviour of $v^z_{x,y }(\omega)$ vs. $\hbar\omega$ could be understood by looking
at $\mu^{azbc}(\omega)$ and $\xi^{aa}(\omega)$ (upper panels). For instance, the structure of
the large peak which leads to a negative $v^z_y (\omega)$ from $~1.4$ to $~1.8$ eV comes
directly from $\mu^{yzxx}(\omega)$ as $\xi^{xx}(\omega)$ and $\xi^{yy}(\omega)$ have several
featureless flat plateaus in that energy region. On the other hand, the
structures of  $v^z_x (\omega)$and $v^z_y (\omega)$ between 2.0 and 2.2\,eV and around  2.7\,eV
come  from the interplay of the $\mu^{azbc}(\omega)$ and both $\xi^{xx}(\omega)$ and
$\xi^{yy}(\omega)$, as prescribed by Eqs.~\eqref{e.z21} and \eqref{e.z22} at
$\alpha=\pi/4$.

\begin{figure}[]
\centering
\includegraphics[scale=.55]{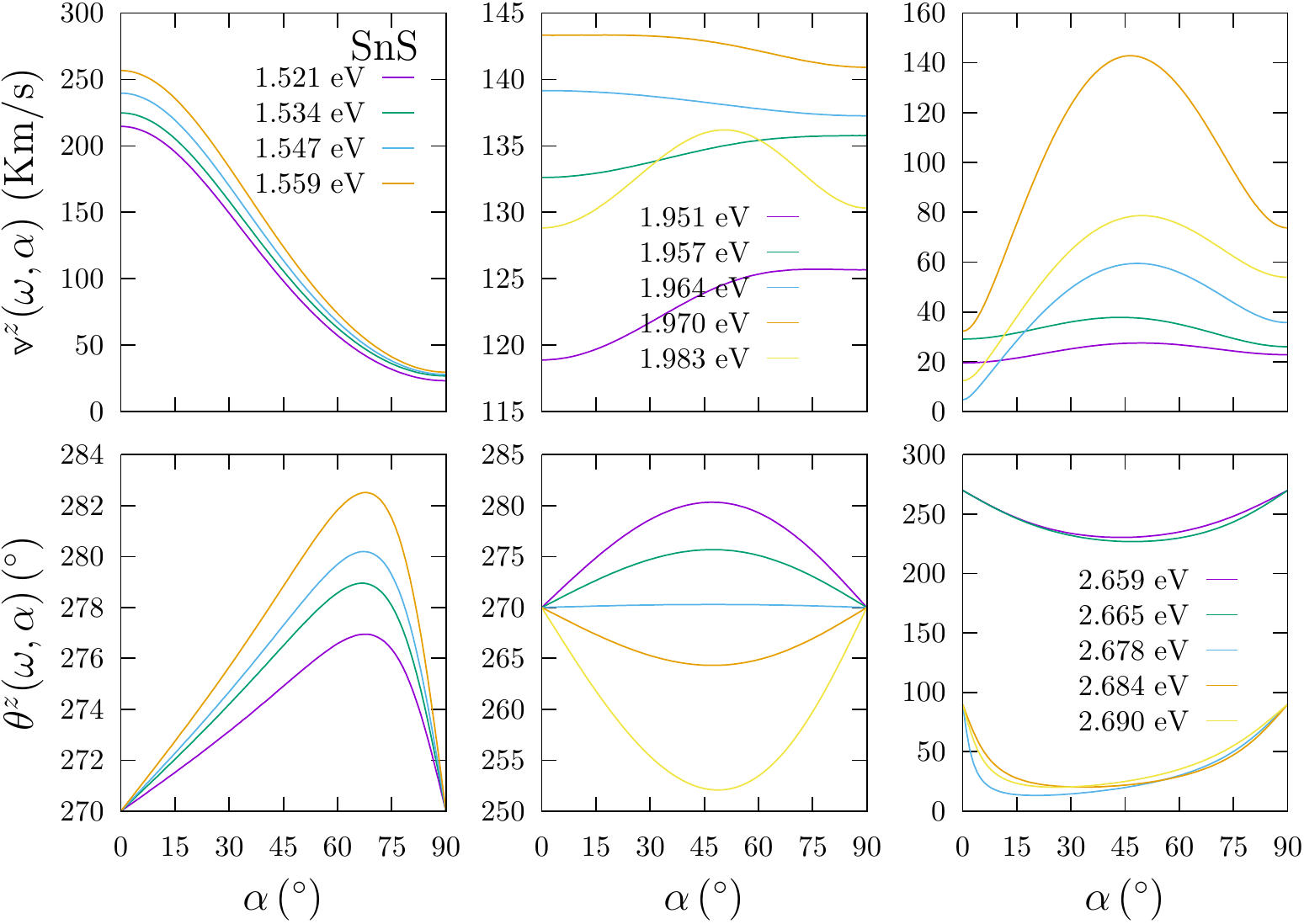}
\caption{(Color Online)
$\mathbb{v}^z(\omega,\alpha)$
 and $\theta^z(\omega,\alpha)$ of Eqs.~\eqref{e.z23} and
\eqref{e.z24}, respectively, vs. $\alpha$ for SnS.}
\label{fig:all-sns}
\end{figure}

We analize the dependence of $\mathbb{v}^{z}(\omega,\alpha)$ and $\theta^z(\omega,\alpha)$
(Eqs.~\eqref{e.z23} and \eqref{e.z24}), with $\alpha$, the angle that gives the
direction of the linearly polarized electric field of the normally incident
beam of light. To do so, we zoom into the three regions of $\hbar\omega$
mentioned above, for which we find not only a large value of the SVI but also
an interesting behavior that allows the manipulation of the direction of the
SVI itself trough the value of $\alpha$. In Fig.~\ref{fig:all-sns} (left panel)
we show $\mathbb{v}^{z}(\omega,\alpha)$ and $\theta^z(\omega,\alpha)$ vs. $\alpha$, for
$\hbar\omega=1.521,\,1.534,\,1.547$ and 1.559 eV,
corresponding to the near infrared. We notice that for
$\alpha=0^\circ$,
$\mathbb{v}^{z}(\omega,\alpha)$
is maximum for the chosen values of $\hbar\omega$ and reaches speeds
around 225 Km/s. As we move towards $\alpha=90^\circ$,
$\mathbb{v}^{z}(\omega,\alpha)$
decreases smoothly and is close to 45 Km/s. Also, we see that at
$\alpha=0^\circ$ and $90^\circ$, $\theta^z(\omega,\alpha)=270^\circ$,
which can be easily
understood as for these angles $\mathbb{v}^{x,z}(\omega,\alpha)=0$, and
thus the spin velocity would be pointing towards $-y$, for an incoming
polarization along $x$ or $y$.  Moreover, we notice that for the other values
of $\alpha$ the spin velocity does not deviate much from the $-y$ direction. In
the middle panel of Fig.~\ref{fig:all-sns},  we show again
$\mathbb{v}^z(\omega,\alpha)$
and $\theta^z(\omega,\alpha)$ vs. $\alpha$, but for
$\hbar\omega=1.951,\,1.957,\,1.964,\,1.970$ and 1.983 eV, corresponding to the
Red subrange of the visible. For these choices of $\hbar\omega$ we see that the
variation of
$\mathbb{v}^z(\omega,\alpha)$
vs. $\alpha$ is smaller  than from the
previous cases and is non monotonic as a function of $\hbar\omega$. On the
other hand the behavior of $\theta^z(\omega,\alpha)$ is monotonic for the new set of
chosen values of $\hbar\omega$, and it display a very interesting behavior for
$\hbar\omega=1.964$\,eV for which
$\theta^z(\omega,\alpha)=270^\circ$, i.e. is constant as a function of $\alpha$;  on
top of this, we notice that the corresponding
$\mathbb{v}^z(\omega,\alpha)\sim 137$ Km/s, is also almost constant as a function of
$\alpha$.
This means that regardless of the orientation of the polarization of
the impinging electric field, the spin-velocity current will be directed along
$-y$ with nearly a constant speed;
we find a similar behaviour for
SnSe (see the appendix). Therefore, there are energies of
the incoming light for which one can inject $z$-polarized spins along the $y$
surface direction regardless of the polarization angle $\alpha$,
thus opening the possibility of having SVI for
unpolarized light.

Above behaviour is easily explained from Fig.~\ref{fig:mu-chi-vaz-sns}, where
for $\hbar\omega=1.964$\,eV, we
see that as $\mu^{yzxx}(\omega)\sim\mu^{yzyy}(\omega)<0$, $\xi^{xx}(\omega)\sim\xi^{yy}(\omega)$, and
$\mu^{xzxy}(\omega)\sim 0$, then Eq.~\eqref{e.z22} will imply that
$\mathbb{v}^{yz}(\omega,\alpha)$
is almost independent of $\alpha$, and  Eq.~\eqref{e.z21}
implies that
$\mathbb{v}^{xz}(\omega,\alpha)\sim 0$.
Therefore, Eq.~\eqref{e.z24} gives
$\alpha=270^\circ$.
Finally, in Fig.~\ref{fig:all-sns} (right panel), for
$\hbar\omega=2.659,\,2.665,\,2.678,\,2.684$ and
2.690 eV, in the Blue region of the spectrum, we show again
$\mathbb{v}^z(\omega,\alpha)$
and
$\theta^z(\omega,\alpha)$ vs. $\alpha$. For
$\mathbb{v}^z(\omega,\alpha)$
we see a monotonic behaviour as a function of $\hbar\omega$, where
$\alpha=45^\circ$ maximizes the speed that could be as large as $\sim140$ Km/s.
 Now, what is interesting to see is that  the  resulting SVI shows similar
values of $\theta^z(\omega,\alpha)$ in the third quadrant  for energies around
(2.659,2.665)\,eV  and in the first quadrant for energies around
(2.678,2.690)\,eV.

Finally, we calculate  the average distance $d$ by which the up and down  spin
populations are displaced.
From  Ref.~\onlinecite{bhatPRL05} we  obtain that
$d\sim 4\tau \mathbb{v}^z(\omega,\alpha)$,
 where $\tau$ is the momentum relaxation time. For instance,
assuming $\tau=100$ fs,\cite{hubnerPRL03} from Fig.~\ref{fig:mu-chi-vaz-sns} we
see that $\mathbb{v}^z(\omega,\alpha=0)\sim 260$\,Km/s for $\hbar\omega=2.151$\,eV and
then $d\sim 104$\,nm. This value of $d$ is $\sim 5$ times larger than those
experimentally  measured values of $d=20$ nm for GaAs/AlGaAs\cite{stevensPRL03}
and $d=24$ nm for ZnSe,\cite{hubnerPRL03}
making the SnS, SnSe, GeS and GeSe monochalcogenides, excellent
structures for the realization of PSC.

\section{Conclusions}
\label{sec:conclusions}
Using novel single-layer 2D monochalcogenides, SnS, SnSe, GeS, and GeSe, we
have shown that these 2D films are excellent candidates for spin current
injection. In particular, we reported the results of \emph{ab initio}
calculations for the spin velocity injection (SVI) due to one-photon
absorption of the linearly polarized light as a function of its energy and
direction of polarization. The theoretical formalism  to calculate the SVI
includes the excited coherent superposition of the spin-split conduction bands
that arise in the noncentrosymmetric structures considered here. We made the
calculations for the cases when the spin is polarized in the $z$ direction that
is perpendicular to the 2D films and study the resulting SVI velocity along
the $x$ or $y$ directions parallel to the films. We have shown that the  SVI
display an  interesting behavior, which depends upon the material, but where we
found similarities in that all of them show values around and above 150\,Km/s
for the SVI, with an ample control of the direction of the SVI through the
manipulation of the angle $\alpha$ of the linearly polarized light. Also,  the
particularities of each material made these structures excellent spintonic
candidates.
 In particular we found that for  SnS and SnSe there are energies of
 the incoming light for which one can inject $z$-polarized spins along the $y$
surface direction regardless of the polarization angle $\alpha$  of the
linearly polarized light, thus opening the possibility of having SVI for
unpolarized light.

The speed values obtained here are of the same order of magnitude as those of
Ref.~\onlinecite{najmaiePRB03}  in unbiased semiconductor quantum well
structures, and  Ref.~\onlinecite{zapataPRB17} in hydrogenated graphene
structures,  while they are an order of magnitude
higher compared to 3D bulk materials. Moreover, the distance $d$ by which the
spin up and spin down populations are separated is larger
than for other semiconductors where $d$ has been
measured.\cite{stevensPRL03,hubnerPRL03} Therefore, the 2D monochalcogenide
structures
considered here are  excellent candidates for the development of spintronics
devices that require pure spin current (PSC).


\section{Acknowledgment}
L.J.R. acknowledges  support by CONACyT through a postdoctoral research
fellowship. B.S.M. acknowledges  the support from CONACyT through
grant A1-S-9410. B.M.F. thank NSF DMR-2015639 and NERSC-DOE-AC02-05CH11231.

\appendix

\section{Results for SnSe, GeS and GeSe}
We go over the results for SnSe, GeS and GeSe, which qualitatively are very
similar to those presented for SnS
in the main text. However, we point out only the most relevant features, since
the detailed explanation could be done following that of SnS.
\begin{widetext}
\begin{center}
\begin{figure}[]
\centering
\includegraphics[scale=1.2]{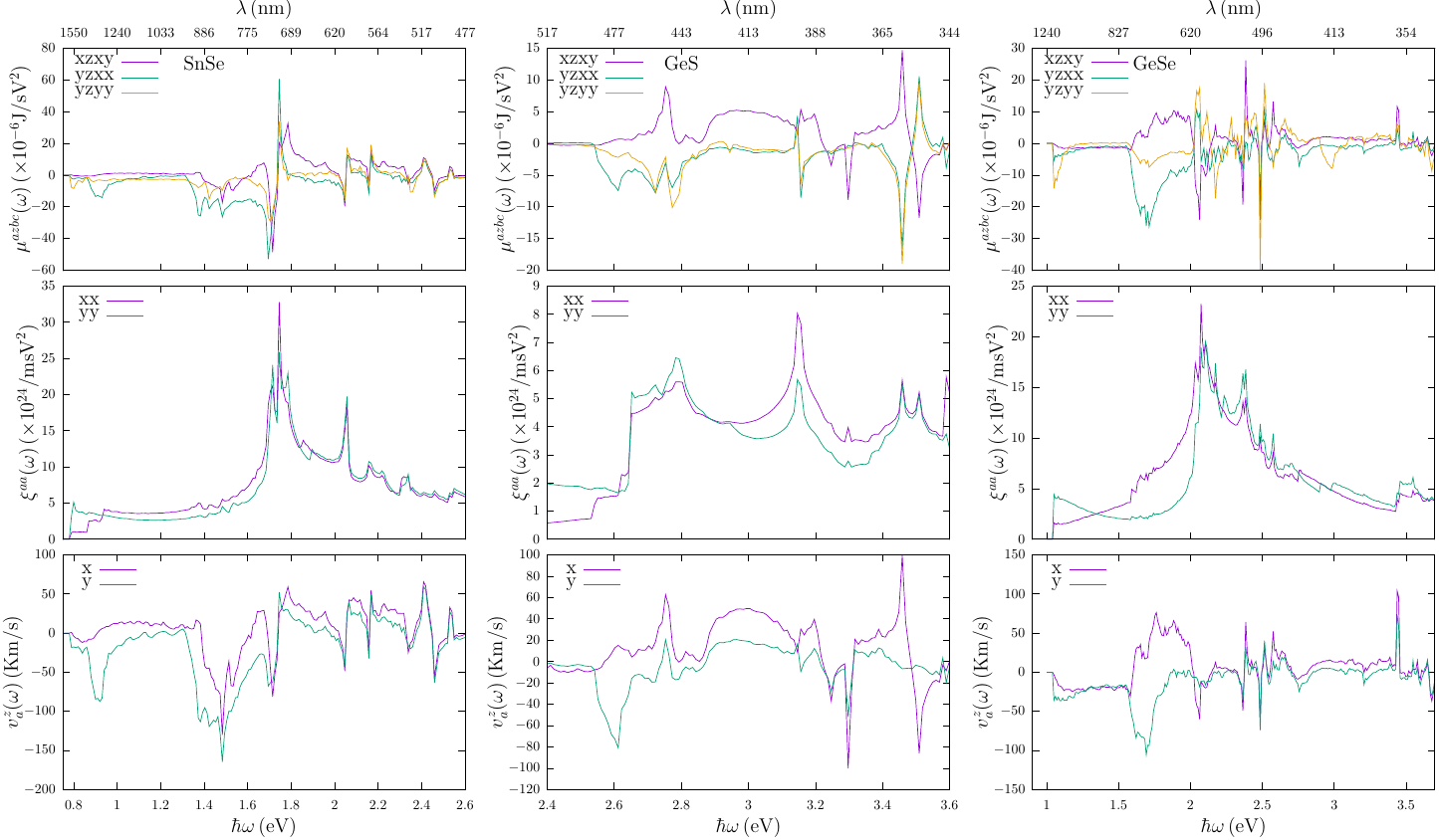}
\caption{(Color Online)
$\mu^{azbc}(\omega)$, $\xi^{aa}(\omega)$, and  $v^z_{(x,y)}(\omega)$  of Eqs.~\ref{eq:mu}, \ref{eq:xi},
\ref{e.z25}  respectively, vs $\hbar\omega$ for SnSe, GeS and GeSe.}
\label{fig:all-mu-chi-vaz-crop}
\end{figure}
\end{center}
\end{widetext}

In Fig.~\ref{fig:all-mu-chi-vaz-crop}
we show  as a function $\hbar\omega$,
for SnSe,  GeS,
and
GeSe,
$\mu^{azbc}(\omega)$, $\xi^{aa}(\omega)$, and $v^z_{(x,y)}(\omega)$ of Eqs.~\ref{eq:mu}, \ref{eq:xi}, and
Eq. \ref{e.z25}, of the main text, respectively. We see  that $v^z_{x,y}(\omega)$, for
each structure has an interesting  behaviour as a function of $\hbar\omega$,
reaching maximum values around $-150$, $\pm 100$, and $-100$ Km/s for SnSe, GeS
and GeSe, respectively, each at different regions of $\hbar\omega$.

\begin{figure}[]
\centering
\includegraphics[scale=.55]{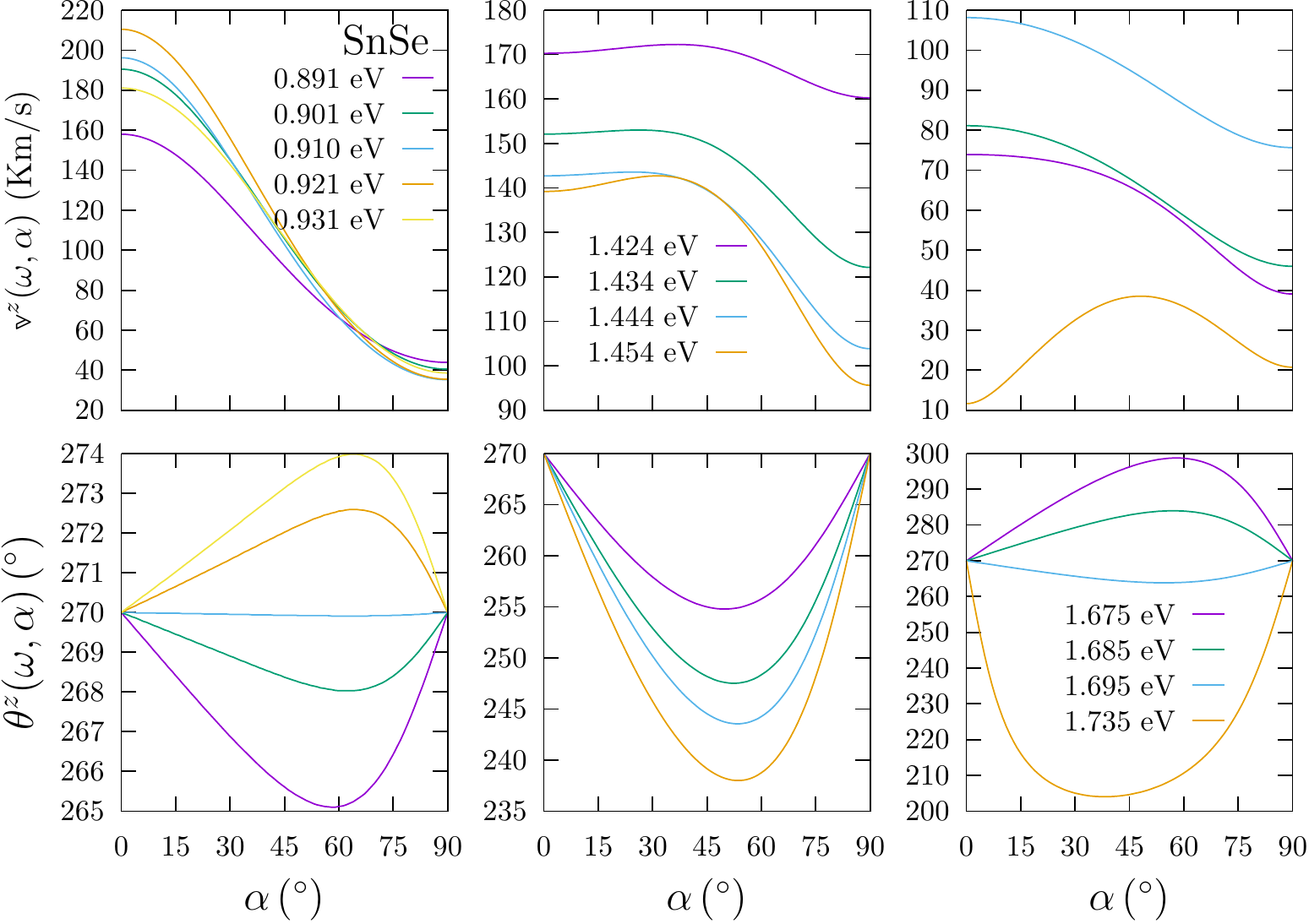}
\caption{(Color Online)
$\mathbb{v}^{z}(\omega,\alpha)$ and $\theta^z(\omega,\alpha)$ of Eqs.~\ref{e.z23} and
\ref{e.z24}, respectively, vs. $\alpha$ for SnSe. }
\label{fig:all-snse}
\end{figure}
\begin{figure}[]
\centering
\includegraphics[scale=.55]{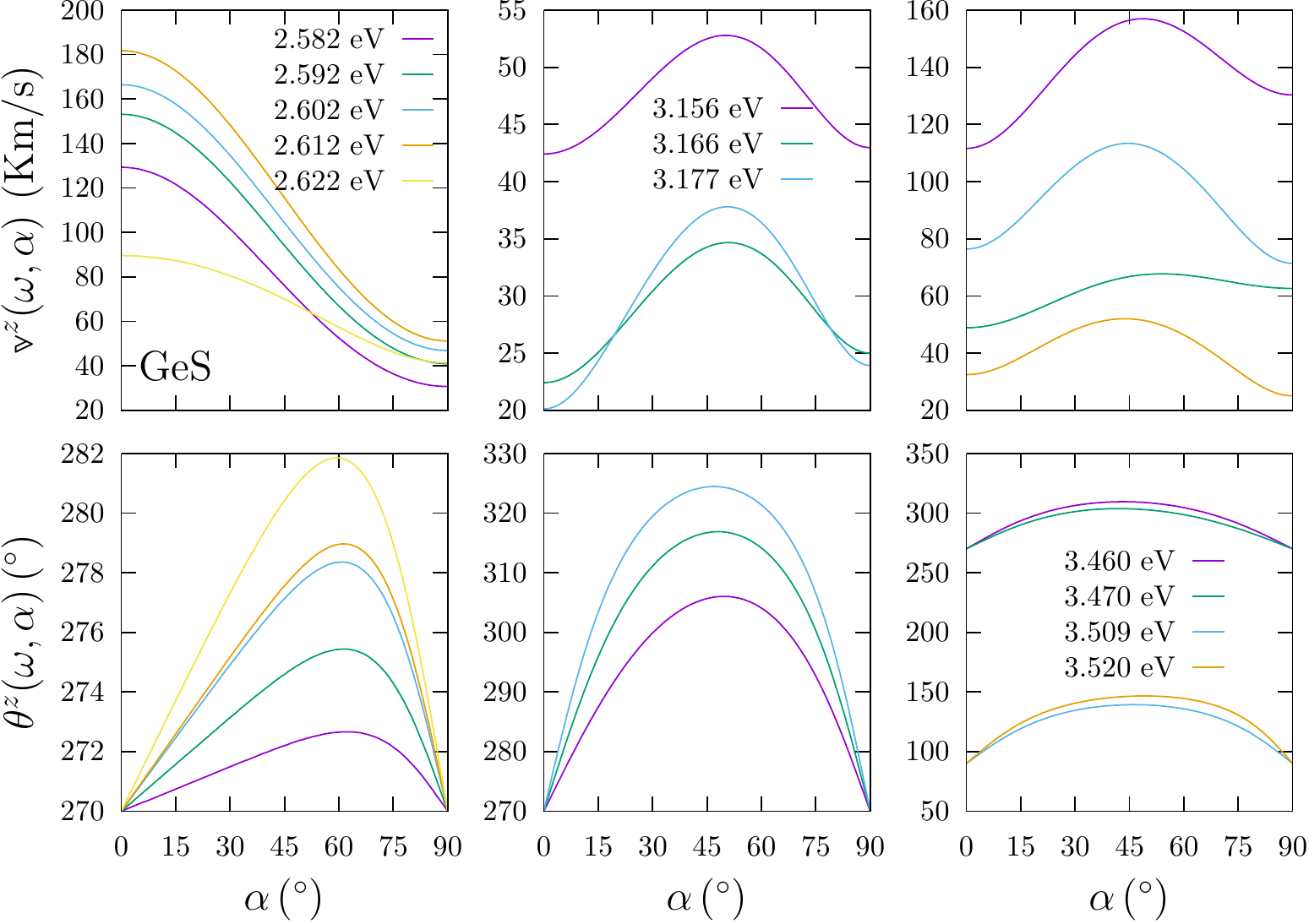}
\caption{(Color Online)
$\mathbb{v}^{z}(\omega,\alpha)$ and $\theta^z(\omega,\alpha)$ of Eqs.~\eqref{e.z23} and
\eqref{e.z24}, respectively, vs. $\alpha$ for GeS. }
\label{fig:all-ges}
\end{figure}

As for SnS, we identify three energy regions for each structure where
$v^z_{x,y}(\omega)$ are large. For SnSe the $\hbar\omega$ regions are  [0.891,0.931]
and $[1.424,1.454]$\,eV, both in the near infrared, and $[1.675,1.735]$\,eV in
the Red subregion of the visible spectrum.
Then, for GeS we have $\hbar\omega\sim [2.582,2.622]$\,eV in the blue,
$[3.156,3.177]$\,eV  in the violet, and $[3.460,3.520]$\,eV in the near UV.
Finally, for GeSe we have $\hbar\omega\sim [1.713,1.733]$\,eV in the Red,
$[1.984,2.064]$\,eV  in the Orange, and $[2.485,2.576]$\,eV in the Blue. As
explained for SnS, the features seen in $v^z_{(x,y)}(\omega)$ readily come from the
interplay of $\mu^{azbc}(\omega)$ and $\xi^{aa}(\omega)$ shown in the upper panels of the
corresponding figure for each system, from where they can be deduced in detail.
\begin{figure}[]
\centering
\includegraphics[scale=.55]{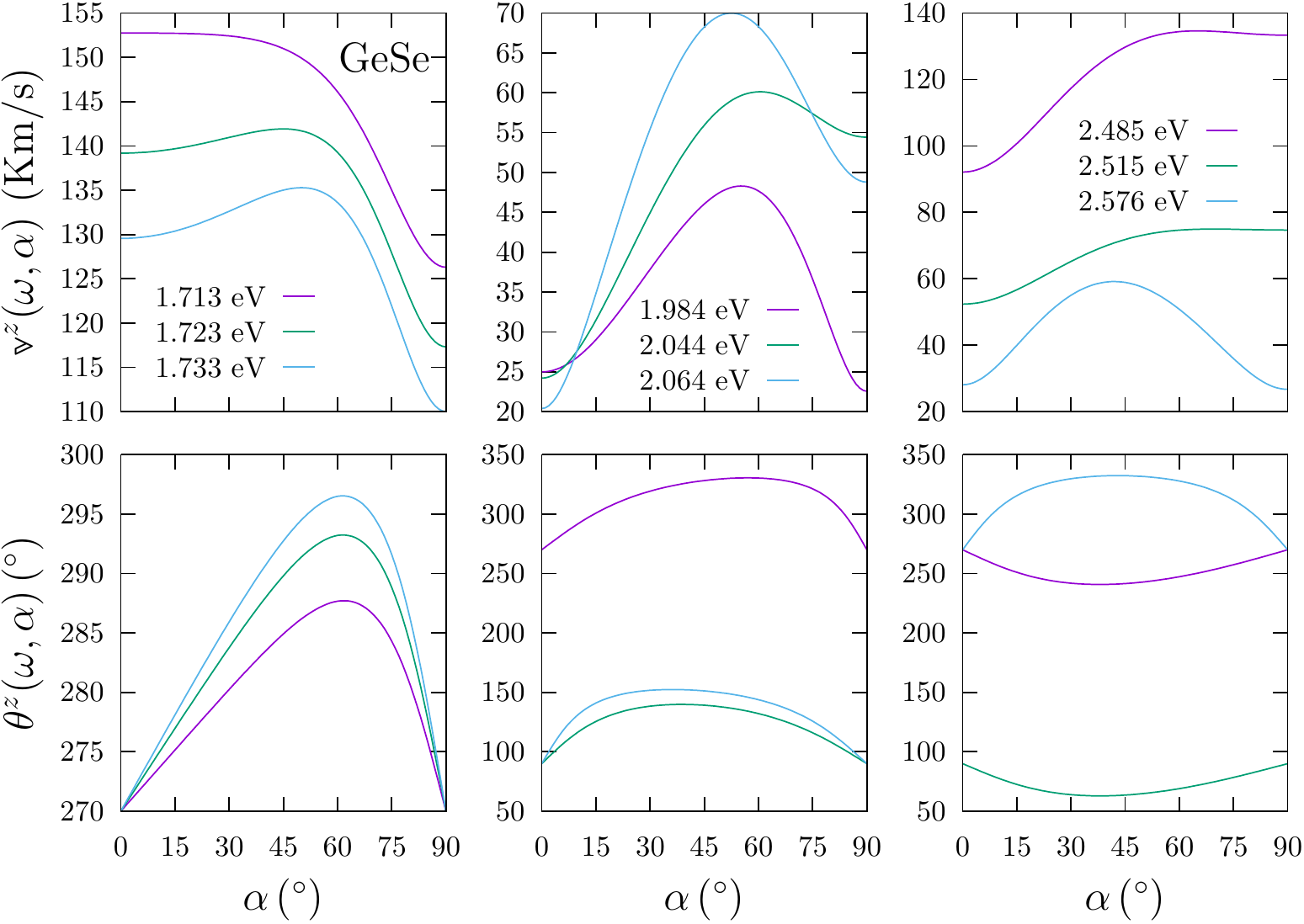}
\caption{(Color Online)
$\mathbb{v}^z(\omega,\alpha)$ and $\theta^z(\omega,\alpha)$  of Eqs.~\ref{e.z23} and
\ref{e.z24},  respectively, vs. $\alpha$ for GeSe. }
\label{fig:all-gese}
\end{figure}
(Eqs.~\ref{e.z23}  and \ref{e.z24} of the main text),
vs. $\alpha$, which is the angle that gives the direction  of the linearly
polarized electric field of the normally incident beam  of light,  also shows a
very interesting behaviour for SnSe, GeS and GeSe, as shown in Figs.
\ref{fig:all-snse} for SnSe, \ref{fig:all-ges} for GeS and
\ref{fig:all-gese} for GeSe, where each panel corresponds to the energy
regions given in the previous paragraph  for each system. We find very large
values of the SVI and also an interesting behavior that allows the manipulation
of the direction of the SVI itself trough the value of $\alpha$, just as
explained in the main text for SnS.
For instance we see that for SnSe at
0.910\,eV (Fig.~\ref{fig:all-mu-chi-vaz-crop}), the direction of the resulting SVI
$\mathbb{v}^z(\omega,\alpha)$ (Eq.~\eqref{e.z25} of the main text) is
$\theta^z(\omega,\alpha)=270^\circ$ independent of $\alpha$, just as we obtained for SnS at
1.964\,eV and shown in Fig.~\ref{fig:mu-chi-vaz-sns} of the main text. Also,
for SnSe we see that for $\hbar\omega=1.424$\,eV  $\mathbb{v}^z(\omega,\alpha)$ is
almost constant with a value of ~165\,Km/s, with a variation of $\theta^z(\omega,\alpha)$
close to the $-y$ direction. Going through the results of GeS and GeSe we can
find similar behavior, and in general we see that the single-layer 2D
monochalcogenides SnS, SnSe, GeS, and GeSe, offer a wide set of possibilities
to manipulate with  linearly polarized light the  spin-velocity injection (SVI)
of a pure spin current (PSC).

\bibliographystyle{apsrev}
\bibliography{ref}

\end{document}